\documentclass[smallabstract,smallcaptions]{dccpaper}

\usepackage{epsfig}
\usepackage{citesort}
\usepackage{amsmath}
\usepackage{amssymb}
\usepackage{color}
\usepackage{url}
\usepackage{graphicx}
\usepackage{epstopdf}
\usepackage{multirow}
\usepackage{amsmath}
\usepackage{amssymb}
\usepackage{float}
\usepackage{subfigure}
\usepackage{booktabs}
\usepackage[justification=centering]{caption}

\newlength{\figurewidth}
\newlength{\smallfigurewidth}

\begin{document}

\title
{\large
\textbf{Varifocal Multiview Images: Capturing and Visual Tasks}
}

\author{%
Kejun Wu${^{1,2}}$, Qiong Liu${^{1,2}}$, Guoan Li${^{1,2}}$, Gangyi Jiang$^{3}$ and You Yang${^{1,2}}$\\[0.5em]
{\small\begin{minipage}{\linewidth}\begin{center}
$^{^{1}}$School of Electronic Information and Communications, Huazhong University of Science and Technology, Wuhan, China \\
$^{^{2}}$Wuhan National Laboratory for Optoelectronics, Wuhan, China \\
$^{^{3}}$Faculty of Information Science and Engineering, Ningbo University, Ningbo, China \\
\url{{wukejun, q.liu, liguoan2021, yangyou}@hust.edu.cn}, \url{jianggangyi@nbu.edu.cn}
\end{center}\end{minipage}}
}

\maketitle
\thispagestyle{empty}

\begin{abstract}
Multiview images have flexible field of view (FoV) but inflexible depth of field (DoF). To overcome the limitation of multiview images on visual tasks, in this paper, we present varifocal multiview (VFMV) images with flexible DoF. VFMV images are captured by focusing a scene on distinct depths by varying focal planes, and each view only focused on one single plane.
Therefore, VFMV images contain more information in focal dimension than multiview images, and can provide a rich representation for 3D scene by considering both FoV and DoF. The characteristics of VFMV images are useful for visual tasks to achieve high quality scene representation. Two experiments are conducted to validate the advantages of VFMV images in 4D light field feature detection and 3D reconstruction. Experiment results show that VFMV images can detect more light field features and achieve higher reconstruction quality due to informative focus cues. This work demonstrates that VFMV images have definite advantages over multiview images in visual tasks.
\end{abstract}

\Section{1 Introduction}

Multiview has become increasingly crucial from the recent advances of consumer electronics and industrial devices, such as mobile cameras, plenoptic cameras and industrial vision systems \cite{7780428}. Researches on multiview are flourishing for meeting the immediate requirements of visual servoing or fuzzy systems for industrial manufacturing \cite{8543489,9133495}, visual defect inspection \cite{8949715,9099905}, industrial internet of things (IIoT) \cite{8815938,9447225}, intelligent video surveillance and summarization \cite{8846765,8765236}, interactive and immersive media \cite{9431676,9105988}. These applications benefit from the wide and flexible field of view (FoV) that multiview provides. However, multiview images are focused on single depth with inflexible depth of field (DoF). Generally, optical imaging systems are constricted by DoF due to the nature of optics, yielding partially focused images within a limited DoF \cite{8712708}. The trade-off between large DoF and high signal-to-noise ratio (SNR) of captured images has been extensively investigated \cite{9381234}. To obtain higher SNR in image formation, the lens aperture should be set large as far as possible, which brings about shallow DoF subsequently. On the contrary, excessive DoF shows a fundamental limitation to the SNR, leading to poor image quality. Consequently, the limited DoF is usually extended by focal stack images, which are a set of images that serially focused on multiple depths of a scene \cite{Wu2020GaussianGI}. However, focal stack images are captured in single view with fixed FoV.

Multiview images have flexible FoV but inflexible DoF, while focal stack images are feasible to extend DoF with high quality but unfeasible to extend FoV. Thus, multiview and focal stack systems are inadequate to meet the demand of visual applications as a result of their shortcomings in DoF or FoV. For example, both high spatial resolution and wide focus range are required for defect inspection applications to scan parts with complex geometries, such as a fastener, a hole, or a highly curved filet area. A low f-number of camera is typically adopted to obtain the best resolution, resulting in a narrow DoF \cite{Harding2016MultifocusHR}.  the post-capture refocusing of light field image allows to control DoF after image acquisition \cite{10.1145/2665074}. However, the DoF can only be controlled in a certain view with limited resolutions, generally the center view. Moreover, multiview images can theoretically be recovered from focal stack images by solving inverse problem, but the FoV of scene cannot be extended, or even the recovered image edges are discarded for recovering quality \cite{5539854,Prez2016LightfieldRF}. Thus, wide FoV and large DoF are an immediate requirement for the aforementioned applications, which are suffering the shortcomings of these two types of data.

The human vision system (HVS) adaptively adjust the shape of eye's lens to focus on objects at different depths, and generates images with varying sharpness in retina. The Perception of 3D are from binocular parallax and accommodation. Inspired by the HVS, we combine multiview and focal stack to complement each other by adopting their advantages to overcome disadvantages. In this paper, we propose a new type of data, varifocal multiview (VFMV) images to describe scenes with flexible DoF. The "varifocal" signifies variable camera focus settings; for instance, focal lengths or focal planes. VFMV images are captured by focusing a scene on distinct focal planes for multiple views, and each view only focused on one single plane. Each view achieves high image quality due to the narrow DoF. VFMV images can be conceptually regarded as the integration of multiview images and focal stack images. Such type of data has a unique advantage that captured images can be adjusted to bracket any specific FoV and DoF. The adjustable representation of scenes can be achieved with wide FoV and large DoF, which enhances the capacity of scene understanding in immersive multimedia and extends the feasibility of industrial visual system. For example, accommodation-vergence conflict (VAC) in VR/AR applications hinders visual performance and causes visual fatigue \cite{Xia2019TowardsAS}. VFMV images are expected to resolve VAC by dynamically shifting focal planes at different depths for different views \cite{Chang2018TowardsMD}. Moreover, VFMV images will contribute to in defect inspection, 3D visual sensing \cite{Zhao20193DVS}, autonomous vehicles, computational photography \cite{Pramila2016ExtractingWF,Lee2016RobustAS}, microscopic imaging \cite{Wu:20,8462234}, telemedicine and telehealth \cite{Chang2011EmergingHO,Kumar2018CompressionOC}, industrial video applications \cite{8372240,8998152}.

\begin{figure}[t]
    \centering
    \includegraphics[width=1\linewidth]{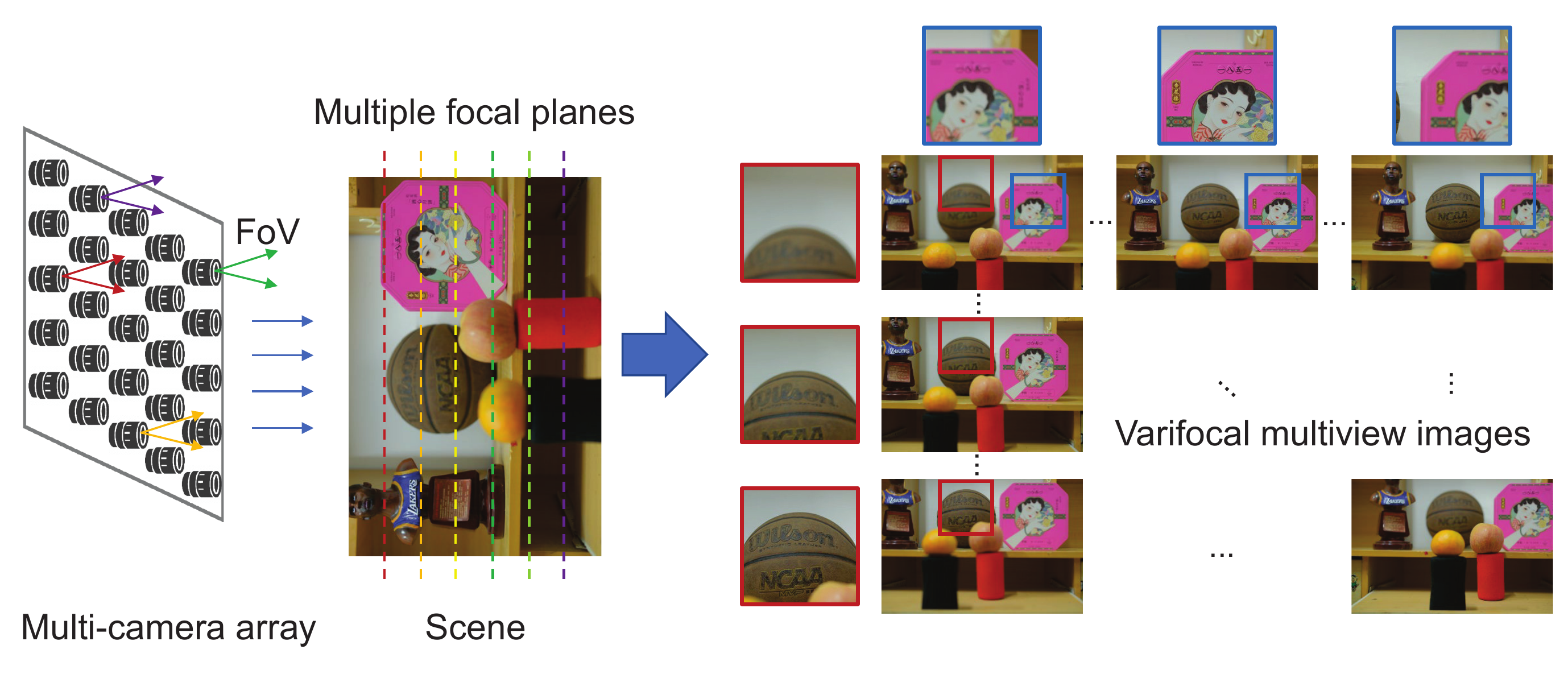}
    \caption{The acquisition of VFMV images. VFMV images can be captured by multi-camera array. The close-ups show that there coexist differences in parallax and focusing degree among views. 
    }
    \label{fig:DataCharacteristic}
    \end{figure}

The rest of this paper is organized as follows. Sec. 2 gives a detailed description of VFMV images. Sec. 3 presents the VFMV system for potential applications. Experiments and discussions are given in Sec. 4. We summarize this paper in Sec. 5.

\Section{2 Data Description}

Multiview or focal stack images are generally subject to invariable DoF or FoV. Considering the limitation and advantage of multiview and focal stack, we propose a new type of multiview, VFMV images with narrow and distinct DoF for each view. The varifocal description refers to a camera lens has a range of focal lengths or a scene has shifting focal planes. It can obtain wide FoV for all the views and high SNR for each view due to narrow DoF. VFMV images have a definite advantage in flexible DoF over conventional multiview images with fixed focal settings. 

The schematic diagram of VFMV acquisition and image characteristics are illustrated in Fig. \ref{fig:DataCharacteristic}. The VFMV acquisition system is shown in the left part of Fig. \ref{fig:DataCharacteristic}.  Multi-camera array of single-camera gantry captures special multiple views focused on diverse focal planes with distinct DoF. It is a specialized system designed for describing objects or scenes in spatial, angular and depth/focal dimensions. The right part of Fig. \ref{fig:DataCharacteristic} shows the image characteristics of VFMV. The focusing degrees of all views are different because each view of VFMV images is focused on distinct focal planes. The parallaxes among views are depended on the camera arrangements. The close-ups show that there exist horizontal and vertical parallaxes and focal differences among views. We can find that VFMV images are highly structured and redundant in spatial, angular and focal dimensions. The distinctive redundancies of VFMV images are not only caused by camera movement but also focused depth varying. Unlike ordinary video sequence where motion vector is competent to describe the inter-view correlations, VFMV needs an additional variable to define and describe the focused degree changes among views. 

\begin{figure}[t]
    \centering
    \includegraphics[width=1\linewidth]{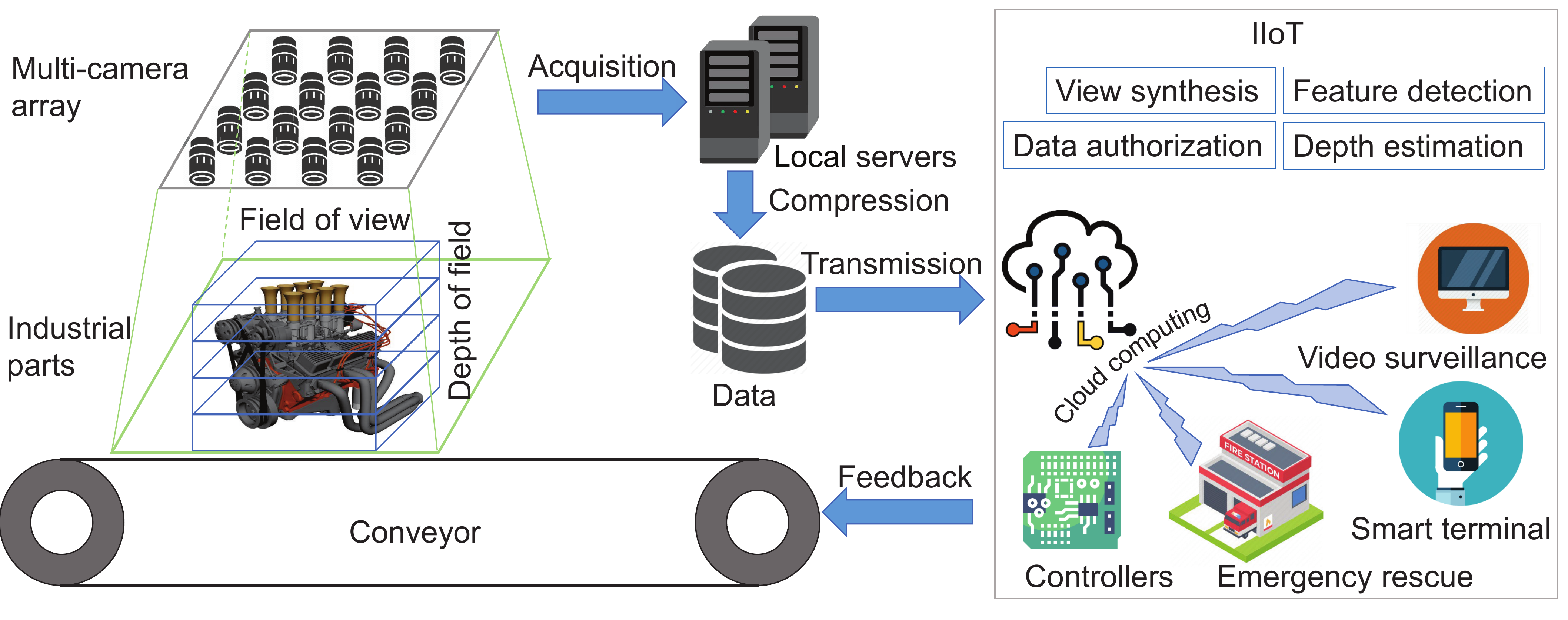}
    \caption{ VFMV system in IIoT among connected devices. 
    }
    \label{fig:SampleScenario}
\end{figure}

Multiview images with high density also refer to as 4D light field images, which are commonly represented the two-plane parameterization $\boldsymbol I(u,v,s,t)$. In the parameterization, $(u,v)$ describes the camera planes, while $(s,t)$ signifies the imaging planes. VFMV images additionally contain focal information. Accordingly, VFMV images can be modeled as a 5D parameterization $\boldsymbol I(u,v,s,t,f)$, where $(f)$ denotes the varying focal lengths or focal planes. Thus, the $(u,v)$, $(s,t)$ and $(f)$ dimensions are the angular dimensions, spatial dimensions and focal dimensions, respectively. The magnification of each views slightly changes due to distinct focal settings. Therefore, image registration is required to align the captured images in each view. We adopt the enhanced correlation coefficient maximization algorithm for the registration \cite{4515873} as follows:

\begin{equation}
    \boldsymbol I(u,v,s^*,t^*,f) = \boldsymbol H \times \boldsymbol I(u,v,s,t,f)
    \label{eq:Registration}
    \end{equation}
where $\boldsymbol I(u,v,s,t,f)$ and $\boldsymbol I(u,v,s^*,t^*,f)$ are the VFMV images before and after registration for a certain view $(u,v)$ and certain focal plane $f$, respectively. $\boldsymbol H$ represents the homography matrix, which is defined as follows:

\begin{equation}
\boldsymbol H=\left[\begin{array}{ll}
\boldsymbol{A} & \boldsymbol{T} \\
\boldsymbol{V} & h
\end{array}\right]
\label{eq:HomographyMatrix}
\end{equation}
where $\boldsymbol{A}$ is the scale, rotation shear components, while $\boldsymbol{T}$ denotes the translation component; $\boldsymbol{V}$ signifies the intersection relation between original line and transformed line; $h$ is the scaling factor.

%VFMV images contain rich scene information, which are capable of widening the applicability of imaging systems and providing strong focus and angular cues for industrial applications. VFMV images can be conceptually regarded as the integration of multiview and focal stack, which makes the best use of their advantages to compensate their disadvantages, achieving adjustable FoV and DoF.

\Section{3 VFMV System}

Thanks to the distinctive image formation and characteristics of VFMV images. Various visual applications in entertainment, medical and industrial fields will benefit from VFMV images. Multiview or focal stack images are generally subject to invariable DoF or FoV.To overcome the limitation of multiview or focal stack systems,  we design the first VFMV system for industrial defect inspection. Fig. \ref{fig:SampleScenario} gives an example of VFMV system for industrial internet of things (IIoT) applications. The system is composed of VFMV acquisition, conveyor and IIoT devices. 
First, multi-camera array or single camera gantry are able to fully cover the industrial workpiece to be defected. The focusing degree of each views are different by changing focal lengths or varying focal planes of cameras. First, VFMV images of the industrial parts are acquired by single shot of camera array or multiple shots of camera gantry with adjustable FoV and DoF. The captured VFMV images are stored in local servers. 
Then, VFMV images are compressed into bitstreams and transmitted to the IIoT. Inside the IIoT, various visual tasks can be conducted; for instance, feature detection, 3D reconstruction, data authorization and depth estimation and so on.

%New view synthesis helps to generate more dense views for a clear description of the scene. 
Feature detection is used to match the multiple views for exploiting the inter-view correlations. 3D reconstruction employs the multiview images in RGB color to render a 3D scene. Depth estimation benefit the 3D representation of the scene with depth information. Besides defect the industrial parts on highly automated production lines, the data of production environment in factory, including fire risk, intrusion and collision warning are also allowed to be collected and processed. On the other hand, the data authorization and trustworthiness are not to be neglected, especially considering high quality and low latency requirements. Data are shared among connected devices in the IIoT, including video surveillance, phone applications, fire station and controllers of the conveyor system. The controllers give feedback to conveyor system for suspending, brake or accelerate the conveyor. Consequently, industrial control of high quality and high precision can be realized by considering adjustable FoV and DoF of VFMV images.

\Section{4 Experiments}
This section conducts two experiments to validate the advantage of VFMV images over multiview images in visual tasks, such as 4D light field feature detection and 3D reconstruction. 

\begin{figure}[t]
    \centering
    \includegraphics[width=1\linewidth]{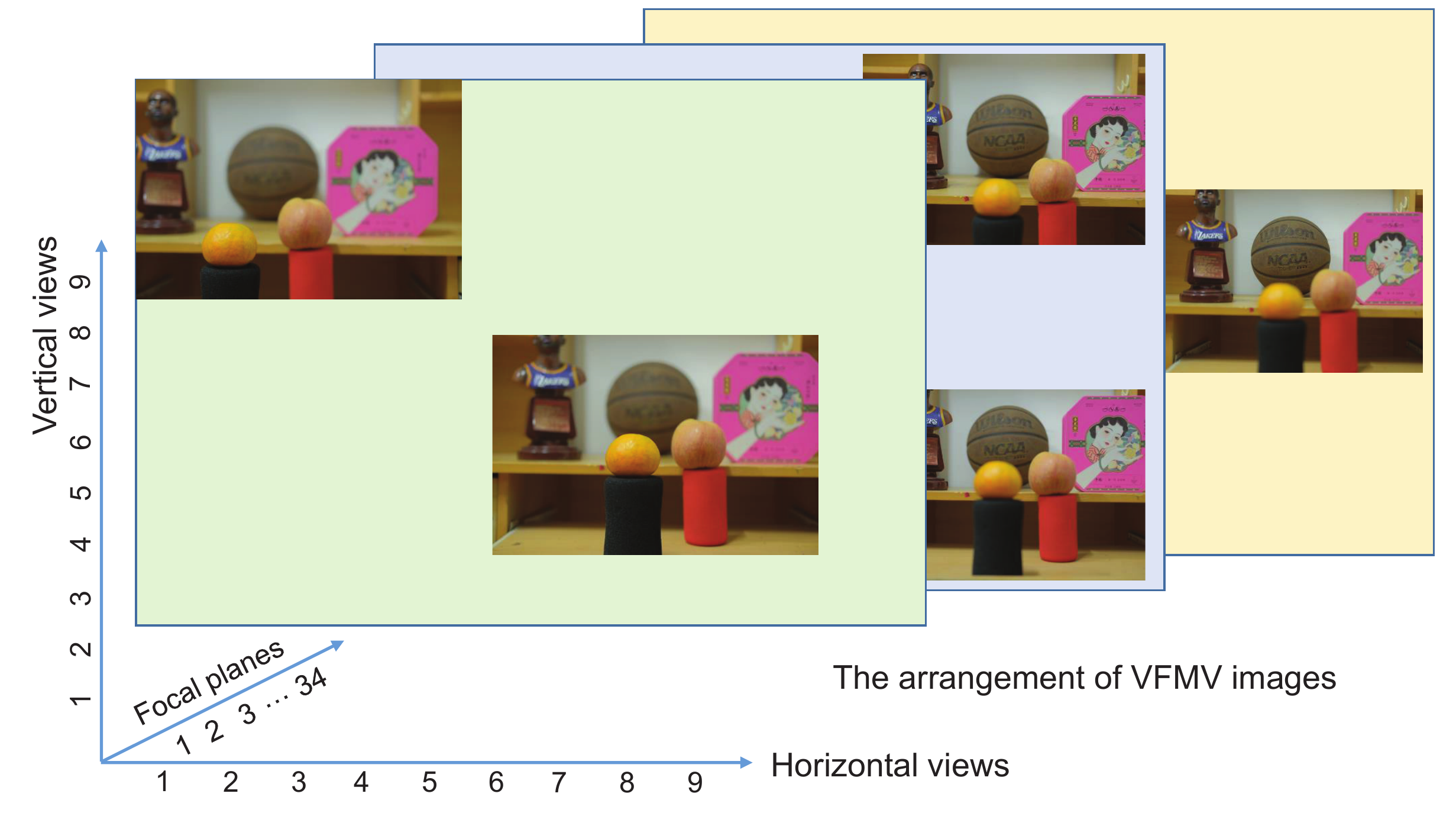}
    \caption{ The arrangement of VFMV images in angular, spatial and focal dimensions.
    }
    \label{fig:Thumbnails}
\end{figure}

\SubSection{4.1 Experiment settings}
In these experiments, we construct an acquisition platform by mounting a Fuji X-S10 digital camera on a X-Y macro focusing rails. It allows consistent adjustments in millimeters when shifting the camera along the X or Y-axis of rails. Thus, VFMV images can be captured by moving the camera position and varying the focal planes simultaneously. We capture scene "Statue" to generate VFMV images and multiview images with $9\times9$ views and resolution $768\times512$. Specifically, there are 34 focal planes to be selected from near to far for all the views, and each view of VFMV images is only focused on a certain focal planes. Multiview images are uniformly focus on the first focal plane from the 34 focal planes. The thumbnails of captured VFMV images used in our experiments are shown in Fig. \ref{fig:Thumbnails}. We can find that the captured VFMV images have $9\times9$ angular arrangement. Views have different in-focus regions because they are focused on different planes in focal dimension.

\begin{figure}[htbp]
    \centering
    \subfigure[]{
    \includegraphics[width=0.46\linewidth]{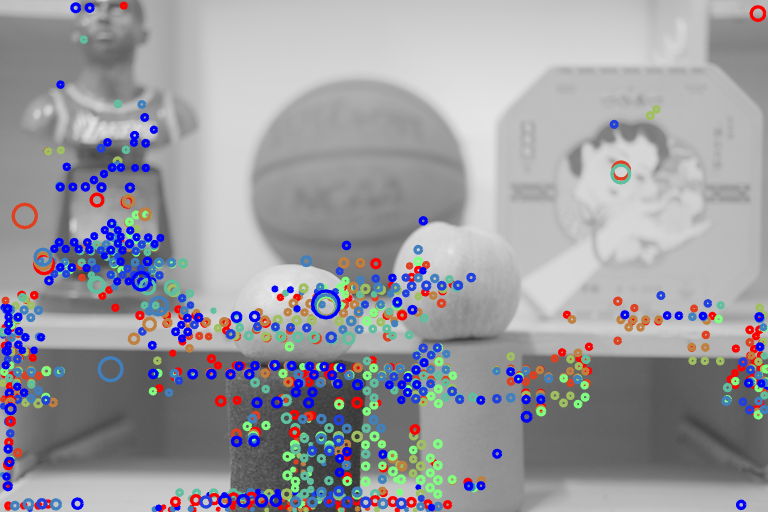}
    }
    \subfigure[]{
    \includegraphics[width=0.46\linewidth]{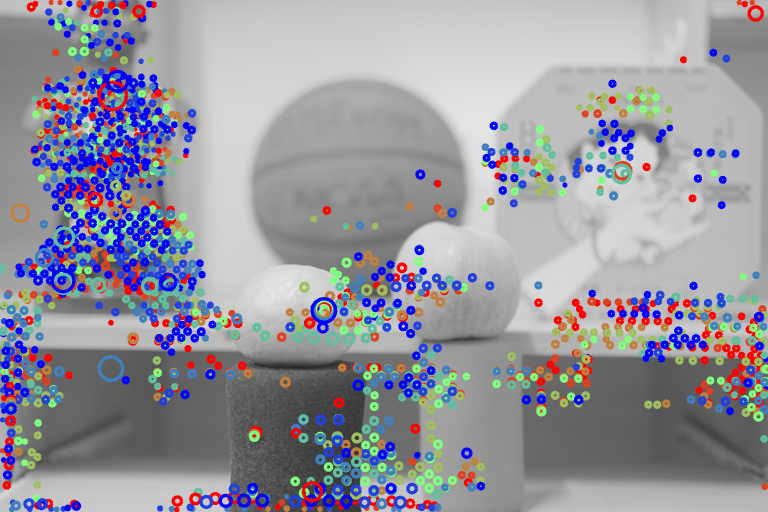}
    }
    \caption{The detected light field features plotted on the central view. (a) Multiview with 1218 features. (b) VFMV with 2737 features. }
    \label{fig:LightFieldFeature}
\end{figure}

\SubSection{4.2 Light Field Feature}
Light field features (LIFF) is a scale invariant features with well-defined scale and depth \cite{8953686}. The informative features are robust to changes in perspective. Thus, it will benefit certain visual tasks where matching features across views play vital roll. In this experiment, LIFF is employed to assess the information richness of VFMV and multiview. LIFF makes full use of all the $9\times9$ views to detect 4D feature\footnote{https://github.com/doda42/LiFF}. The configurations are as follows: peak detection threshold of 0.0015, non-edge selection threshold of 10, 4 octaves begin from -1, 3 levels per octave. The light field features are plotted on the central view shown as Fig. \ref{fig:LightFieldFeature}. 
In Fig.~\ref{fig:LightFieldFeature} (a), 1218 features are detected in multiview images, while 2737 features for VFMV images. It illustrates that VFMV images contain more rich information than multiview images. It is also worth noting that the features of VFMV images are with more wide distributions and accurate locations than that of multiview images. VFMV images deliver more informative features, because views of VFMV images are focused on distinct focal planes, which provides adjustable DoF and focus cues. The focus cues of VFMV images enhance the ability to detect light field feature. 

\SubSection{4.3 3D Reconstruction}

In this experiment, 3D reconstruction is conducted by two views structure from motion. Structure from motion is one of the vital technologies to implement vision-based 3D reconstruction from a collection of images, which contributes to geosciences, localization and navigation systems and cultural heritage. We run the open source project of 3D reconstruction\footnote{https://github.com/yihui-he/3D-reconstruction}. SIFT feature is performed to detect points of interest in structure from motion. The matched features are further employed to find intrinsic matrix and estimate rotation and translation vectors. The point pairs are finally plotted on 3D mesh. We capture the scene "Statue" to generate VFMV images and multiview images, both of which contain two views. The left and right views of multiview images are focused on the middle depth of the scene, while the left and right views of VFMV images are focused on middle and middle-back depth, respectively.
 
% LightFieldFeature

\begin{figure}[htbp]
    \centering
    \subfigure[]{
    \includegraphics[width=0.46\linewidth]{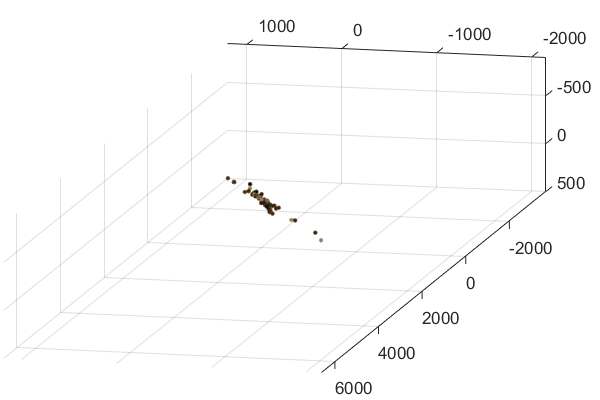}
    }
    \subfigure[]{
    \includegraphics[width=0.46\linewidth]{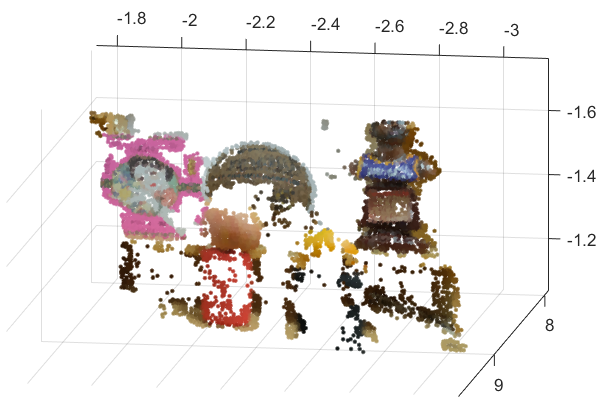}
    }
    \caption{3D reconstruction. (a) The reconstructed result of multiview images are incomplete. (b) The reconstructed result of VFMV images are overall clear regardless of some area flaws. }
    \label{fig:SFM}
\end{figure}

The reconstruction results are shown in Fig.~\ref{fig:SFM}. We can find that VFMV images achieve higher reconstruction quality than multiview images. In Fig.~\ref{fig:SFM} (a), the reconstructed scene from multiview images are generally incomplete, because the input multiview images are focused at single depth with limited DoF information, which yields less matched feature. In Fig.~\ref{fig:SFM} (b), the reconstructed scene from VFMV images are overall clear regardless of some area flaws. This is because VFMV images contains distinct focal information and conveys focus cues. The focus cues provides informative DoF and enhances the image reconstruction quality.

\Section{5 Conclusions}
In this paper, we propose an idea of VFMV images to meet the requirements of multiview in entertainment and industrial visual tasks. VFMV images contain more information in focal dimension, and can provide a rich representation for 3D scene by considering both FoV and DoF. The nature of VFMV images is useful for visual tasks to achieve high quality scene representation. It overcomes the limitation of multiview images on visual tasks due to flexible FoV but inflexible DoF. Experiments on light field feature detection and 3D reconstruction show that VFMV images achieve high performance than multiview images in visual tasks.

\Section{Acknowledgment}
This work was supported in part by the National Key Research and Development Program of China under Grant 2020YFB2103501, in part by the National Natural Science Foundation of China under Grant 61971203, in part by the Wuhan Science and Technology Bureau under Grant 2020020601012222.

\Section{References}
\bibliographystyle{IEEEbib}
\bibliography{refs}

\end{document}